# Photonic hook: A new curved light beam

Liyang Yue,[1,*] Oleg V. Minin,[2] Zengbo Wang,[1] James N. Monks,[1] Alexander S. Shalin,[3] Igor V. Minin[4,*]

[1]*School of Electronic Engineering, Bangor University, Bangor, LL57 1UT, UK*
[2]*National Research Tomsk State University, Lenin Ave., 36, Tomsk, 634050, Russia*
[3]*ITMO University, 49 Kronverksky Ave., 197101, St. Petersburg, Russia*
[4]*National Research Tomsk Polytechnic University, Lenin Ave., 30, Tomsk, 634050, Russia*
*\*Corresponding authors: l.yue@bangor.ac.uk and prof.minin@gmail.com*

**Abstract**
It is well-known that electromagnetic radiation propagates along a straight line, but this common sense was broken by the artificial curved light - Airy beam. In this paper, we demonstrate a new type of curved light beam besides Airy beam, so called 'photonic hook'. This photonic hook is a curved high-intensity focus by a dielectric trapezoid particle illuminated by a plane wave. The difference of the phase velocity and the interference of the waves inside the particle cause the phenomenon of focus bending.

The Airy waveform was first theorized as a solution to the Schrödinger equation by Berry and Balázs, et al. in 1979 [1]. In 2007 Siviloglou et al. successfully observed an Airy beam in both one-/two-dimensional (1D/2D) configurations [2]. As an Airy beam propagates, it will bend to form a parabolic arc in a medium, rather than diffract due to its uncommon beam profile (an area of principal intensity and less luminous areas to infinity) [2]. Unique properties of the Airy beam, e.g. free acceleration [3, 4], self-healing [5], and large field of view when forming a light sheet [6], have been reported in the generally acknowledged scientific literature, and afforded potentially important applications for imaging and manipulating microscale objects [6, 7]. Meanwhile, a microparticle can act as a refractive micro-lens to focus the light wave within a subwavelength volume [8, 9]. This generates a narrow, highly intensive, weak-diverging beam known as 'photonic jet', which propagates into the background medium from the shadow side of a particle [10, 11].

To the best of our knowledge, there is no research published about the formation of curved photonic jets. In this paper, we report a new possibility to produce the curved light beam, named 'photonic hook' to emphasise its curvature characteristic [12]. The key difference between the photonic hook and the Airy beam is that the photonic hook is created using the focusing of a plane wave through an asymmetric dielectric-particle which is the combination of a wedge prism and a cuboid. This process should be simpler than generation of an Airy beam using a laser and a spatial light modulator [2]. Formation of the photonic jet within the teraherz band in a symmetrically cuboid particle has been investigated in our previous publication [12]. Relying on the established knowledge, the influence of size and prism angle of an asymmetric particle on the distribution of field enhancement, width (full width of half maximum, FWHM), and curvature of the photonic hook are effectively quantified in this study.

The numerical model of a symmetric/asymmetric particle irradiated by a plane wave is built in the commercial finite integral technique (FIT) software package – CST Microwave Studio. A 532-nm wavelength ($\lambda$) plane wave propagates from $+z$ to $-z$ direction, and is polarised along the $y$ axis. The model represents a dielectric cuboid with the addition of a wedge-shaped prism made of the same material on the face encountering plane wave, while the same sized cuboid without prism is also simulated as a reference. The refractive indices of the particle material (transparent fused silica) and background medium (air) are set up to 1.46 and 1.0 respectively [13]. The cuboid structure has equal length, width and height. Side length, $L$, is normalised to $\lambda$, and angle $\alpha$ defines the inclined plane of the prism.

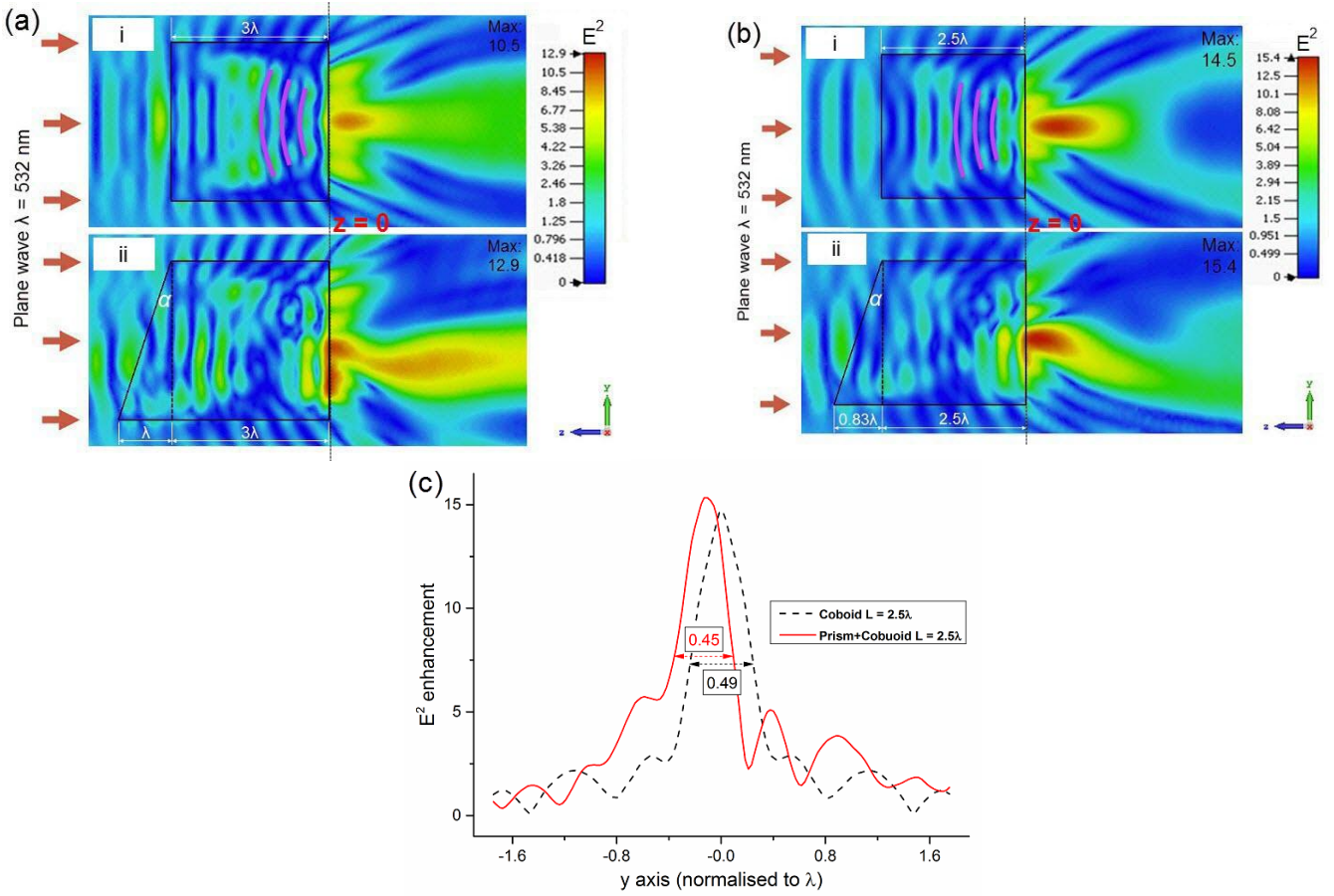

Fig. 1. $E^2$ field intensity distribution for (a) $L = 3\lambda$ and (b) $L = 2.5\lambda$. (c) $E^2$ enhancement profiles along the $y$ axis for the symmetric and asymmetric particles with $L = 2.5\lambda$.

Fig. 1 (a) describes the $E^2$ field enhancement distribution of the plane wave passing through the (i) symmetric and (ii) asymmetric particles with $L = 3\lambda$. In Fig. 1 (a) – i, the wave front of a plane wave becomes concave and convergent in the symmetric cuboid particle (pink auxiliary lines), leading to the formation of a typical photonic jet in the background medium. An asymmetric particle assembled by a cuboid of the same size and a wedge prism with $\alpha = 18.43°$ (arctan (1/3)) is illuminated by a plane wave, as shown in Fig. 1 (a) – ii. Here, the prism not only lets the photonic jet shift away from the optical axis, but also bends it in the central propagation, forming a 'hook' shape. The peak $E^2$ enhancement of photonic hook reaches 12.9 in Fig. 1 (a) - ii, which is larger than the 10.4 achieved by the jet from a normal cuboid particle exhibited in Fig. 1 (a) - i. The same 'photonic hook' effect also appears in the asymmetric particle of different size, as shown in Fig. 1 (b) – ii, where $L = 2.5\lambda$ and $\alpha = 18.43°$ (arctan (1/3)). In this case, intensity enhancement of the photonic hook is larger than that achieved in Fig. 1 (a) - ii, reaching 15.4, and the high-intensity area is more concentrated around the bottom surface of particle (right side), as shown in Fig. 1 (b) - ii. Fig. 1 (c) expresses the $E^2$ enhancement profiles along the polarization direction ($y$ axis direction) for the positions delivering maximum enhancements in the models of the symmetric and asymmetric particles with $L = 2.5\lambda$, demonstrated in Fig. 1 (b) ($z = 0$ is set to the flat end of particle, as shown in Fig. 1 (a) and (b), position $z = -289.7$ for symmetric model, and position $z = -112.4$ for asymmetric model). It is shown that FWHM of photonic hook (solid red curve) is shorter than that for photonic jet induced by the symmetric cuboid particle (dashed black curve), and is able to break the simplified diffraction limit (0.5$\lambda$ criterion).

Curvature of a photonic hook is defined by the factor $\beta$ aided by a midline $L_{jet}$, and is the angle between the two lines linking the start point with the inflection point, and the inflection point with the end point of photonic hook, respectively, as shown in Fig. 2 (a). To quantify the influence of the prism shape, we build several models including a $L = 3\lambda$ cuboid coupled with a variety of prisms with different $\alpha$ values. Fig. 2 (b) summarises the hook curvature – $\beta$ (degree, black curve – square markers) and the increase of peak field enhancement (%, red curve – circle markers) compared to those in the symmetric cuboid case.

The $E^2$ distributions at several key $\alpha$ values are illustrated in the insets in Fig. 2 (b) to reveal the curvature changes of the photonic hooks. Symmetric cuboid represented as $\alpha = 0°$, creates 0° for factor $\beta$ and 0% for increase of peak field enhancement in Fig. 2 (b), and a typical photonic jet from the cuboid particle can be seen in inset 1. When angle $\alpha$ is increased from 0° to 4.76°, inset 2, the symmetry of jet becomes broken, and jet starts leaning to the direction of the longer side of asymmetric particle (-$y$ direction in our model), resulting in a 7° $\beta$ factor. The largest curvature, $\beta = 35°$, is achieved at $\alpha = 18.43°$, as shown in inset 3. However, further increase of angle $\alpha$ leads to the decrease of the curvature $\beta$ factor, until almost $\beta = 0°$ at $\alpha = 33.69°$. Inset 4 indicates the regression of $\beta$ at a large $\alpha$ angle, and the corresponding intensity enhancement distribution is characterised as a non-curved jet leaning towards the –$y$ direction, where the start point, the inflection point, and the end point of jet form an almost straight line. In addition, increase of peak intensity enhancement exists in all asymmetric models, and maximum 28.85% increase is obtained at $\alpha = 22.62°$.

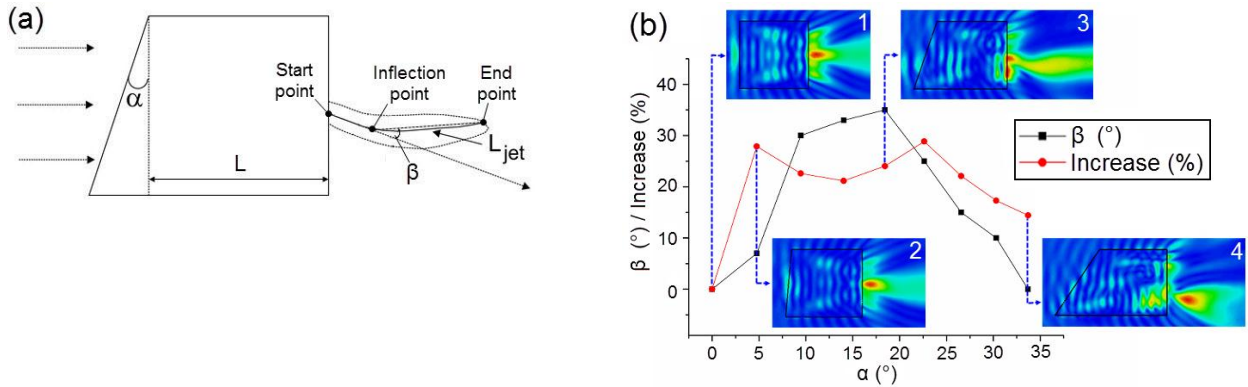

Fig. 2. (a) Diagram of photonic hook (b) Curvatures and intensity enhancements for different $\alpha$.

Traditionally, the generation of strongly confined photonic jet depends on the trade-off between the refractive index contrast (dielectric particle to background media) and the scaling effect of the dielectric particle (relative-to-wavelength size), rather than the particle shape [12]. Consequently, focusing of light by particles with the high degree of spatial symmetry, e.g. spheres, cylinders or disks, is easier to calculate based on the aforementioned conditions and classic Lorenz-Mie theory. However, formation of a photonic jet through a cuboid particle is due to the phase delay along the wave front rather than the factors that govern the focusing by spherical or cylindrical particle [12]. It is represented as a concave and convergent wave front in the cuboid, shown as the pink auxiliary lines in Fig. 1 (a) - i and (b) - i. In the asymmetric particles, upper and lower halves of the particle (+$y$ and -$y$ direction in Fig. 1 (a) - ii and (b) - ii) have different thickness along $z$ axis. The diversity of the thickness with an asymmetric particle produces an unequal phase of the transmitted plane wave which is perpendicular to $y$ axis, resulting in the irregularly concave deformation of the wave front inside particle, leading to the formation of a photonic hook. A light beam with astigmatism is a similar case in the macro scale to help to understand this process [14]. Additionally, in inset 4 in Fig. 2 (b), the large prism angle $\alpha$ provides great contrast of the concave deformation between the upper and lower halves of asymmetric particle. The low-thickness upper half of the particle is more similar to the normal cuboid and may not be able to affect and 'correct' the jet back to the direction of the incident plane wave. However, this capability could be kept in the scenarios of a lower $\alpha$ angle, as shown in insets 2 and 3 in Fig. 2 (b).

Power flow distributions in the asymmetric particles with $\alpha = 18.43°$ and their vicinity are plotted in Fig. 3 (a-d) to analyse the mechanism of the photonic hook formation with the variation of sizes: $L = 1\lambda$, $2.5\lambda$, $3\lambda$, and $4.5\lambda$, respectively. Vortexes in the power flow are stable focuses in the phase space [15], and the power flow couples to the other planes through the singular points at the centre of vortex [16]. In Fig. 3, we can find a transition of 'bending' characteristic for the high-intensity area. Fig. 3 (a) $L = 1\lambda$ has very similar features with the model of symmetric cuboid particle, characterised as a centred non-curved hot spot with symmetric streamlines. Curvature of the hot spot and asymmetry of streamlines gradually appear in the power flow distributions of particles with characteristic sizes from $2.5\lambda$ to $4\lambda$, as shown in Fig. 3 (b) and (c). The longest and most defined photonic hook is found in Fig. 3 (c) and produced by the particle with $L = 3\lambda$. If enlarge the particle to $L = 4.5\lambda$, the hot spot will maintain a long shape, as shown in Fig. 3 (d).

Based on the above power flow analysis, it is shown that photonic hook is the curved area with strong field enhancement, as shown in Fig. 3. Most of the power flow distributions of the normally straight photonic jets only match the profile of high-intensity area in the region where power flows just exit the particle [17]. However, the photonic hook reported in this paper can entirely match the trend of power flows in a certain shadow region. Interference caused by the diversity of particle thickness along the polarization direction produces some more singular points in the vicinity of the asymmetric particle, which induces the curvatures appearing at the streamlines. Asymmetric particles that are smaller than $L = 2.5\lambda$ do not have enough volume to trigger the enormous difference in the phase velocity between the upper and lower halves of the particle, thus cannot create the photonic hook. In addition, it is shown that the location of photonic hook moves away from the shadow boundary of the particle with the increasing size of particle in Fig. 3. In the case of large particle, e.g. Fig. 3 (d), the corresponding long transmission distance of the jet in background medium weakens the influence of phase delay and interference. This produces a high intensity stripe instead of a photonic hook. It could be noted that this hybrid property of the beam profile described above makes it suitable for the optical trapping applications in mesoscale [18] and nanoscale [19-22] for the material and life science research.

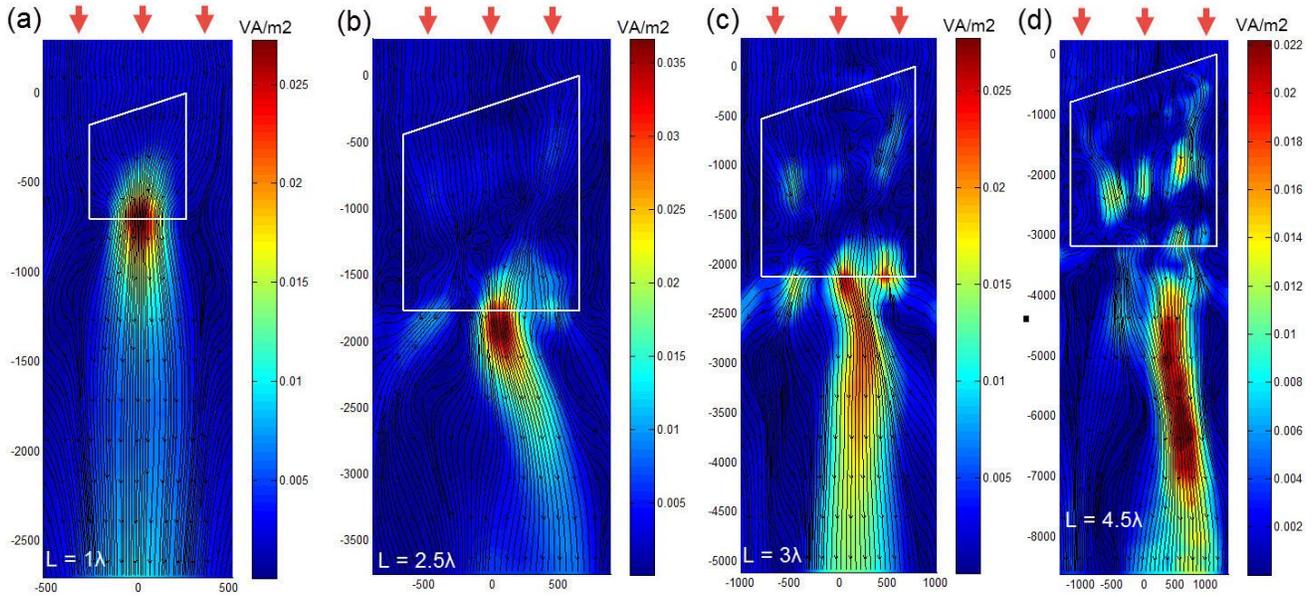

Fig. 3. Power flow diagrams for asymmetric particles $L = 1\lambda$ (a), $2.5\lambda$ (b), $3\lambda$ (c) and $4.5\lambda$ (d) with $\alpha = 18.43°$.

In conclusion, the new type of curved light beam called "photonic hook" is achieved via the numerical simulation of the focusing by an asymmetric dielectric particle, which is a cuboid topped with a wedge prism. It is shown that higher intensity and narrower focus can be offered by this photonic hook compared to those of the jet produced by normal cuboid lens. The power flow analysis of scale models suggests that the photonic hook phenomenon is due to the unequal phase across the particle caused by the diversity of particle thickness and the interference of local fields.

**Funding.** The Sêr Cymru National Research Network in Advanced Engineering and Materials (NRNF66 and NRN113), the Knowledge Economy Skills Scholarships (KESS 2, BUK289), the Mendeleev scientific fund of Tomsk State University (8.2.13.2017), Tomsk Polytechnic University Competitiveness Enhancement Program grant (TPU CEP_INDT_76\2017), Russian Science Foundation Grants (16-12-10287 and 17-79-20051), Russian Fund for Basic Research (16-52-00112), the President Scholarship (SP-4248.2016.1), and the Ministry of Education and Science of the Russian Federation in the frame of GOSZADANIE Grant (3.4982.2017/6.7).